\documentclass[12pt]{article}
\usepackage{graphicx}

\newcommand\pubnumber{}
\newcommand\pubdate{\today}

\def\napoli{Florida International University, Miami, FL 33199, USA}
\def\support{\footnote{Work supported by the US Department of Energy}}

\def\Title#1{\begin{center} {\Large #1 } \end{center}}
\def\Author#1{\begin{center}{ \sc #1} \end{center}}
\def\Address#1{\begin{center}{ \it #1} \end{center}}

\newcommand\pubblock{\rightline{\begin{tabular}{l} \pubnumber\\
         \pubdate  \end{tabular}}}
\newenvironment{Abstract}{\begin{quotation}  }{\end{quotation}}
\newenvironment{Presented}{\begin{quotation} \begin{center} 
             PRESENTED AT\end{center}\bigskip 
      \begin{center}\begin{large}}{\end{large}\end{center} \end{quotation}}

\def\beq{\begin{equation}}
\def\eeq#1{\label{#1}\end{equation}}
\def\eeqn{\end{equation}}

\def\beqa{\begin{eqnarray}}
\def\eeqa#1{\label{#1}\end{eqnarray}}
\def\eeqan{\end{eqnarray}}

\let\bar=\overbar

\def\Dslash{\not{\hbox{\kern-4pt $D$}}}
\def\dslash{\not{\hbox{\kern-2pt $\del$}}}

\def\msb{{\bar{\ssstyle M \kern -1pt S}}}

\begin{document}
\begin{titlepage}
\pubblock

\vfill
\Title{New Results on Three-Nucleon Short-Range Correlations}
\vfill
\Author{ Misak Sargsian\support}
\Address{\napoli}
\vfill
\begin{Abstract}
The recent progress in studies of two-nucleon~(2N) short-range correlations~(SRCs) are reviewed with 
the main emphasize given to the observation of the strong dominance of proton-neutron~(pn) SRCs in nuclei 
as compared to proton-proton and neutron-neutron SRCs. Based on the pn SRC dominance a specific prediction is made for 
the dynamical structure of 3N- SRCs for inclusive $A(e,e^\prime)X$ reactions, 
 according to  which  the 3N- SRCs are generated through the sequence of two short range $pn$ interactions. 
This allowed us to predict that the height of the plateau in the ratios of  inclusive eA cross sections to that of  $^3He$
in the 3N-SRC region  is related to the height of the plateau in the 2N-SRC region by a quadratic relation.  The analysis of the available experimental 
data supports validity of such a relation.

\end{Abstract}
\vfill
\begin{Presented}
Intersection of Particle and Nuclear Physics\\
Indian Wells, CA, USA  May 29 --  June 3, 2018
\end{Presented}
\vfill
\end{titlepage}
\def\thefootnote{\fnsymbol{footnote}}
\setcounter{footnote}{0}

\section{Introduction}
 \vspace{-0.2cm}
Three nucleon short-range correlations~(3N-SRCs) are configurations in  nuclei in
which three nucleons interacting at very short distances  produce a single nucleon in the ground state 
wave function  with a very large momentum ($\ge \sim 700$~MeV/c)  balanced by two nucleons with comparable momenta.  
Such configurations  yet to be  probed experimentally.   Due to their  high density and local structure  3N-SRCs  dominate in  
high momentum component of nuclear wave function which is almost universal,  with the nuclear~($A$,$Z$) dependence
factored in the  SRC coefficients, $a_2$, and $a_3$.

The 3N-SRC is  a 
testing ground for  ``beyond the standard nuclear physics" phenomena such as irreducible three-nucleon forces, inelastic transitions in 3N system as well as transition from hadronic to quark degrees of freedom. Their strenght is expected to be proportional to a higher power  of the local nuclear density thus making them an essential factor  in the dynamics of super-dense nuclear matter. For example  the inclusion of  the irreducible 3N-forces  through the intermediate state $\Delta$-isobars significantly  alters equation of state  of high density nuclear matter predicting neutron star masses $\ge 2M_{\odot}$\cite{PanHei:2000}.
 
Until recently the possibility of direct experimental probing of 2N- and 3N-SRCs  considered to be problematic 
due to  requirements of high-momentum transfer nuclear  reactions in  specific kinematical settings that 
render the  scattering cross sections very small~(see Ref.\cite{FS81} and references therein). 
With the  operation of 6~GeV continuous beam electron accelerator at Jefferson Lab~(Jlab) in 1990's, and unprecedented resolution  in the exploration of  nuclear structure 
has been achieved, which made possible the recent strong progress in SRC studies.

 \vspace{-0.4cm}
\section{Recent Progress in 2N-SRC Studies}
 \vspace{-0.2cm}
The first dedicated investigation of  2N-SRCs  in high momentum transfer inclusive electron-nucleus  reactions~($A(e,e^\prime)X$)   resulted in  the observation of a plateau  in the ratios of  inclusive cross sections of heavy nuclei to  the deuteron \cite{FSDS} measured at SLAC with, $Q^2\ge  2$~GeV$^2$ and  $x> 1.5$. Here  $x = {Q^2\over 2m_N q_0}$ with $m_N$ being the nucleon mass and $q_0$ transferred energy to the nucleus (for a nucleus $A$, $0< x < A$). 
The observed plateau, largely insensitive to $Q^2$ and $x$,  with the magnitude proportional to  the parameter $a_{2}(A,Z)$\cite{FS88} which is the probability of finding 2N-SRCs in the ground state of the nucleus A.  These plateaus were confirmed in the measurements of the  inclusive cross section ratios of nuclei A to  $^3$He\cite{Kim1,Kim2},  at similar kinematics with the magnitude of  plateaus taken to be related to the relative probability, ${a_{2}(A,Z)\over a_{2}(^3He)}$.    These, together with  more recent  measurements of the nuclear to the deuteron inclusive cross section ratios\cite{Fomin2011} provided a compelling evidence for the sizable ($\sim 20\%$)  high momentum component of the ground state nuclear wave function  for medium to heavy nuclei originating from 2N-SRCs.

While the above discussed plateaus  provided the first evidence for 2N-SRCs,  the detailed knowledge  of 2N- SRC dynamics required a semi-inclusive experiments in which one  or both  nucleons  from 2N-SRCs  
have been detected. 

Such first experiments were performed at Brookhaven National Laboratory\cite{eip1,eip2} in which both struck and spectator nucleons from 2N-SRCs in high momentum transfer $A(p,ppn)X$ reactions were detected. The theoretical analysis of these experiments  indicated that the probability of finding proton-neutron combination  in 2N-SRCs exceeds by almost a factor of 20 the analogous probabilities for proton-proton and neutron-neutron SRCs\cite{isosrc}.  This  result was subsequently   confirmed in direct semi-inclusive electroproduction reactions at JLab\cite{eip3,eip4} and both are understood as arising from the dominance of the tensor component in the NN interaction at distances  $|r_1 -r_2|  \le 1$~fm \cite{eheppn2,Schiavilla:2006xx}.  The dominance of the 
$pn$ component in 2N-SRCs suggested a new  prediction for  momentum sharing between high momentum protons and neutrons in asymmetric nuclei\cite{newprops} according to which the minority component will dominate  the high momentum component of the nuclear wave function. This prediction was confirmed indirectly in $A(e,e'p)X$ experiments\cite{twofermi} and directly in in $A(e,e'p)X$ and $A(e,e'n)X$ processes in which protons and neutrons from 2N-SRCs have been probed independently\cite{pndirect}.   

In addition to measuring the isospin content of 2N-SRCs, several semi-inclusive experiments\cite{eip2,eip3,Erez18} 
confirmed expected  ``geometrical" picture of 
2N- SRCs of  overlapped  two nucleons  having relative momentum between $250-650$~MeV/c with back-to-back  orientations  
and with moderate center of mass momentum  $\le \sim 150$~MeV/c. There are  several  reviews\cite{srcrev,srcprog,Atti:2015eda,arnps,Hen:2016kwk}  which  have documented extensively the recent progress  in investigation of 2N-SRCs in a wide range of nuclei.

\begin{figure}[htb]
\vspace{-0.4cm}
\centering
\includegraphics[height=6cm,width=9cm]{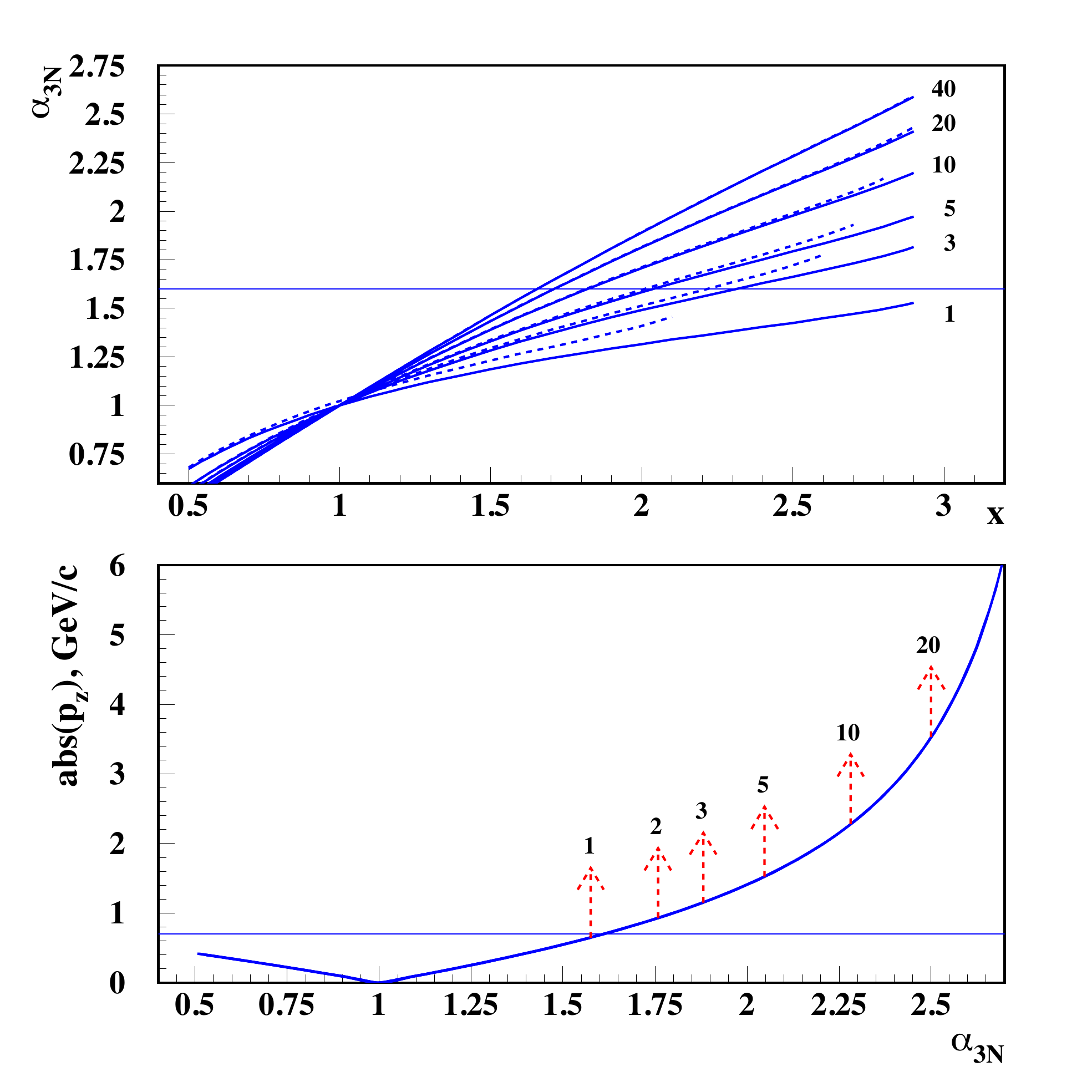}
     \vspace{-0.4cm}
\caption{Kinematics of 3N~SRCs. (upper panel) Relation between $\alpha_{3N}$ and $x$ for $m_S$ calculated  with   $k=0$ (dotted line) and   $k = 250$~MeV/c(dashed line) internal momentum. The curves are labeled by their respective  $Q^2$ values.
(lower panel) The dependence of   $|p_z|$ on $\alpha_{3N}$. Arrows indicate the maximum possible $\alpha_{3N}$'s that can be reached at given values of $Q^2$.}
\label{fig1}
\end{figure}

 \vspace{-0.4cm}
\section {Three Nucleon Short Range Correlations} 

 \vspace{-0.2cm}
 
Despite an impressive progress achieved in studies of 2N-SRCs the confirmation of 3N-SRCs remains  elusive.  
Similar to 2N- SRC,  the one possible  signature of 3N-SRCs is the presence of the {\em plateau}  in the ratio of inclusive cross sections of nuclei A and $^3$He in the kinematic region of $x>2$ - beyond the range dominated by 2N SRCs ($1.5 < x < 2$).
The first observation of such a  plateau at $x>2$ was reported in Ref.\cite{Kim2}. However it was not observed  by subsequent measurements\cite{Fomin2011,Ye:2017mvo}.  

One reason of the absence of the plateau is the modest  $Q^2$ covered in these experiments. 
To determine the minimal $Q^2$ needed to enhance the 3N-SRC contribution in the scattering process one first needed to 
identify the dominant structure of 3N-SRCs in the nuclear ground state.   Our analysis of  three-nucleon systems in 
 Ref.\cite{eheppn2} demonstrated that configurations  in which two recoil nucleons have the smallest possible  mass, $m_S$, dominate in the spectrum of the 3N-SRC nuclear decay function at low excitation energies.  
 This allowed us to conclude\cite{DFSS18} that in  inclusive scattering which integrates over the nuclear excitation energies, the dominant contribution to  3N-SRCs originates from  configuration in which two recoil nucleons produced in the backward direction 
 with respect to the momentum of the interacting nucleon with the mass, $m_S\ge  2m_N$.

With the dominant mechanism of 3N-SRCs established we are able to calculate the kinematic requirements 
best suited for the enhancement  of 3N- SRCs  in inclusive $eA$ scattering.  Due to relativistic nature of SRC kinematics
the most natural description   is achieved through the light-cone~(LC) nuclear spectral functions\cite{FS88,multisrc} in which the correlated nucleons are described by their nuclear  LC momentum fractions, $\alpha_i$ and transverse momenta $p_{i,\perp}$.  
Integrations over the LC kinematics of correlated recoil nucleons and the   transverse momentum  of the interacting nucleon results in a LC density matrix of the nucleus 
$\rho_A(\alpha_N)$ that enters into 
the cross section of inclusive scattering.
Here $\alpha_N$ is the LC momentum fraction of the nucleus carried by the interacting nucleon. 

To  evaluate the LC momentum fraction, $\alpha_{N}$ (denoted henceforth as $\alpha_{3N}$)  
describing the interacting nucleon in  3N-SRC, 
we consider the  quasielastic scattering from  a 3N system:  $q + 3m_N = p_f + p_{S}$, where $q$, $p_f$ and $p_S$ are the four momenta of 
the virtual photon, final struck nucleon and recoil two-nucleon system respectively.  Defining LC momentum fraction, $\alpha_{3N} = 3- \alpha_S$, 
where $\alpha_S = 3{E_S - p_S^z\over E_{3N}-p_{3N}^z}$ in the center of mass of the $\gamma^* (3N)$ system and using the boost invariance of 
the LC  momentum fractions one arrives at the following lab-frame expression (see Ref.\cite{DFSS18} for details) :
\begin{eqnarray}
\alpha_{3N}   & = & 3 \ - \ {q_- + 3 m_N\over 2 m_N} \left[1 \ \  + \ \ {m_S^2 - m_N^2\over W_{3N}^2}  \ \  +  
\right. \ \ \  \nonumber \\  
  & & \left.    
\sqrt{\left(1 - {(m_S + m_N)^2\over W_{3N}^2}\right)\left(1 - {(m_S - m_N)^2\over W_{3N}^2}\right)}\right],
\label{alpha3n}
\vspace{-0.2cm}
\end{eqnarray}
where  $W^2_{3N} = (q + 3m_N)^2 = Q^2{3-x\over x} + 9 m_N^2$ and  $q_- = q_0 - {\bf q}$ with $q_0$ and $q$ being energy and momentum 
transfer in the lab with $z|| {\bf q}$. The above relation allows to identify the kinematical conditions most favorable for the isolation of  3N-SRCs in 
inclusive $A(e,e^\prime)X$ reactions.  This is done by identifying the minimal value of $\alpha_{3N}$ above which one expects 
the  contribution of  3N-SRCs to dominate.  Such a value for $\alpha_{3N}$ was estimated in Ref.\cite{DFSS18} resulting in $\alpha_{3N}^{min} = 1.6$.
With such threshold,  from Eq.(\ref{alpha3n}) we  identify the  most favorable domain in $x$ and $Q^2$ in which to 
search for 3N-SRCs in inclusive $A(e,e^\prime)X$ reactions.  In Fig.\ref{fig1}(a) we present the $\alpha_{3N}$ - $x$ relation for  different values of $Q^2$.  
The figure  shows that starting around  $Q^2\ge 2.5 - 3$~GeV$^2$ 
there  exists a finite kinematic  range with $\alpha_{3N}\ge 1.6$ 
where one expects the onset of the 3N-SRC dominance.  In addition, starting with $Q^2\ge 5$~GeV$^2$ the onset of 3N-SRCs  is practically insensitive to the recoil mass of the spectator system, $m_S$.  
Fig.\ref{fig1}(b) shows the dependence of  longitudinal momentum of the interacting nucleon,  $|p_z|$ on $\alpha_{3N}$ with the arrows indicating the maximum possible 
$\alpha_{3N}$'s that can be probed at given values of $Q^2$. One observes  from the plot that the characteristic momenta of the struck nucleon  in 3N~SRCs for $\alpha_{3N}\ge 1.6$ is $p_{z}\ge \sim 700$~MeV/c. 
 
Another advantage of considering 3N-SRCs in terms of $\alpha_{3N}$,  is that at sufficiently large $Q^2$ the LC momentum 
distribution function $\rho_A(\alpha_{3N})$ is not altered 
due to final state  hadronic interactions~(FSIs). The important feature in high energy limit is that FSI's redistribute  the $p_\perp$ strength in the 
nuclear spectral function leaving $\rho_A(\alpha_{3N})$  practically intact\cite{ms01,BS15,edenx}.   


 \vspace{-0.4cm}

\section{Signatures of 3N-SRCs:}  
 \vspace{-0.2cm}

One of the main properties of  inclusive $eA$ scattering in high $Q^2$ limit is the factorization in the following form: 
\vspace{-0.2cm}
\begin{equation}
\sigma_{eA} \approx \sum\limits_N \sigma_{eN}\rho^{N}_A(\alpha_N),
\vspace{-0.2cm}
\label{eA}
\end{equation}
where $\sigma_{eN}$ is the  elastic  electron-bound nucleon scattering cross section and $\rho^N_A(\alpha_N)$ is the
light-front density matrix of the  nucleus at given LC momentum fraction of the nucleon.

 From the local property of  SRCs  one expects that $\rho^N_A(\alpha_N)$  in the correlation region to be proportional to the light-front density matrix of the two- and three-nucleon systems\cite{FS88,FSDS}. This expectation is the main reason of the prediction of the plateau for the ratios of inclusive cross sections 
 in the SRC region. Similar to 2N-SRCs for the 3N-SRC  one predicts  a plateau for the ratio such as:
\vspace{-0.2cm}
\begin{equation}
R_{3}(A,Z) = {3\sigma_{A}(x,Q^2)\over A \sigma_{^3He}(x,Q^2)}\left|_{\alpha_{3N}>\alpha^0_{3N}}\right.,
\vspace{-0.2cm}
\label{R3}
\end{equation}
 where  $\alpha^0_{3N}$  is the threshold value for the $\alpha_{3N}$ above  which one expects onset of 3N-SRCs (taken as $\sim 1.6$ in our analysis). 

\begin{figure}[!htbp]
\vspace{-0.2cm}
     \centering
 \includegraphics[width=9cm,height=6cm]{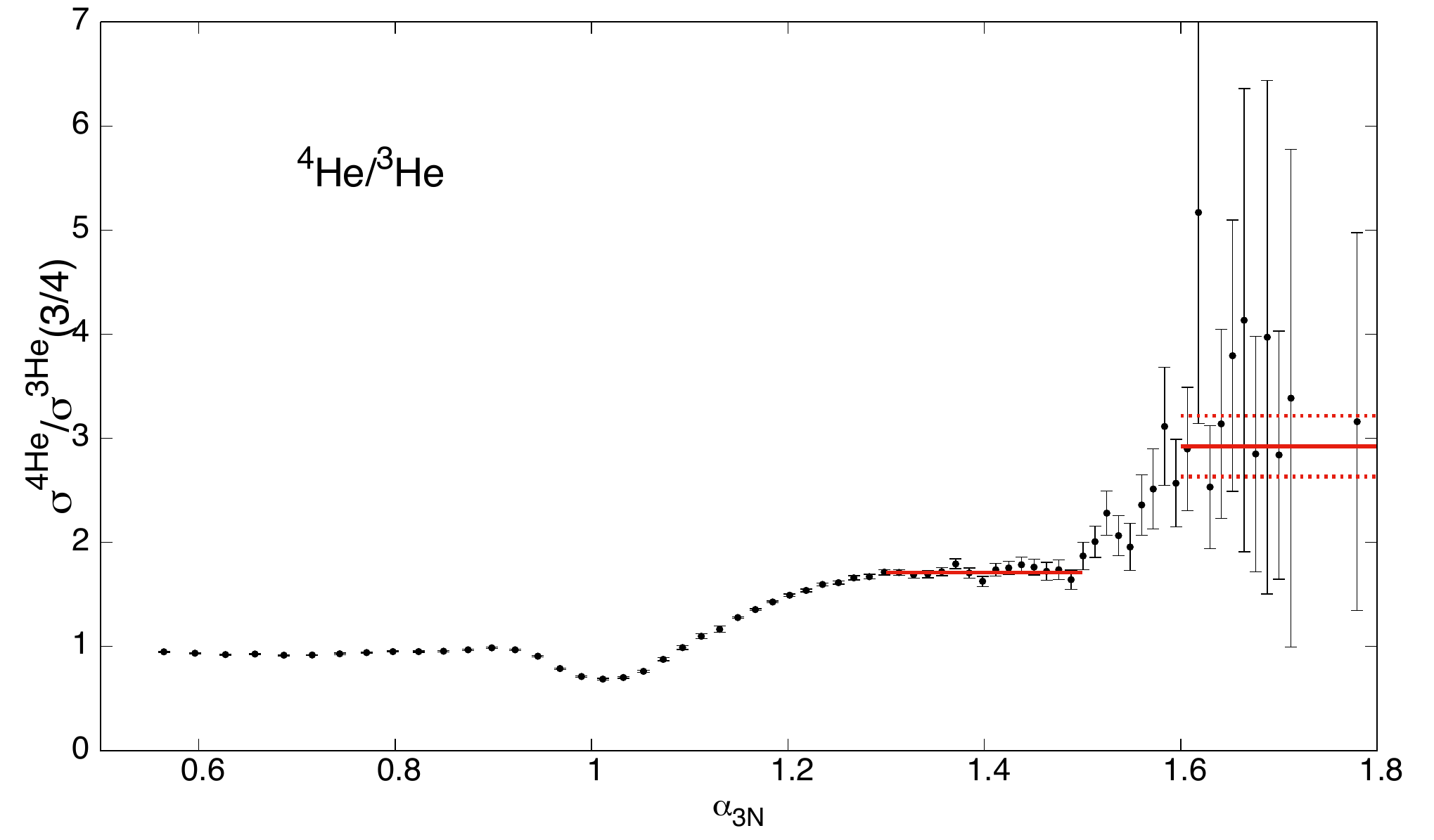}
\vspace{-0.4cm}
   \caption{Per-nucleon cross section ratio for  $^4$He  to $^3$He indicating $a_2(A)\over a_2(^3He)$ in the 2N-SRC regions and the prediction $R_3 = R_2^2$ in the 3N region, $1.6 < \alpha_{3N} < 1.8$. See Eq.~\ref{R3_to_R2sq}.}
      \label{4He_3He_ratio}
   \end{figure}

  In Fig.~\ref{4He_3He_ratio} one finds the ratio of cross sections, ${3\sigma^{4He} \over 4\sigma^{3He}}$ for the 
  largest available $Q^2\sim 2.7$~GeV$^2$ from experiment\cite{Fomin2011} as a function of the $\alpha_{3N}$.
  As figure shows there is a strong indication of the onset of the plateau at $\alpha_{3N}>1.6$ that can be attributed to the onset of 3N- SRCs.
  In this figure one observes also a plateau in the $1.3 \le \alpha_{3N} \le 1.5$ region which is related to the dominance of the 2N-SRCs. It is interesting that the similar pattern is observed also for other nuclei such as  $^9$Be, $^{12}$C, $^{64}$Cu and $^{197}$Au albeit with the larger errors in the $\alpha_{3N}\ge 1.6$ region\cite{DFSS18}.
  
 \vspace{-0.6cm}
  \section{3N-SRCs and the {\boldmath $pn$} Dominance:} 
 \vspace{-0.2cm}

The theoretical analysis of   3N-SRCs which contribute in inclusive scattering\cite{multisrc,DFSS18}  indicated that such correlations are  
produced in the sequence  of  two short-range $NN$ interactions in which the  produced fastest 
nucleon  interacts with the  external probe. 
The presence of  short-range $NN$ interactions  in the 3N-SRC configuration  makes the  
recently found $pn$-SRC dominance\cite{isosrc,eip4,eip3} a profound phenomenon also for  3N-SRCs. 

For 3N-SRCs in this case one expects that only $pnp$ or $npn$ configurations will contribute   with minority component playing role of ``catalyst" 
in forming a fast interacting nucleon with momentum $p_i$.
For example in the case of $pnp$ configuration, the  neutron will play a role of intermediary in furnishing a large momentum transfer to one of the protons 
with two successive  short range $pn$ interactions.  Quantitatively such a scenario is reflected in  $\rho^N_{A(3N)}(\alpha_N)$ which in the domain of 3N-SRC is expressed through $pn$- SRCs as follows:
\vspace{-0.2cm}
 \begin{eqnarray}
 \rho^N_{A(3N)}(\alpha_N,p_\perp)  & \approx &     \sum\limits_{i,j}\int F(\alpha^\prime_i,p_{i\perp},\alpha^\prime_j,p_{j\perp}) \times \ \ \ \ \ \ \ \ \ \ \  \nonumber \\  
& &  \rho^N_{A(pn)}\left(\alpha^\prime_i,p^\prime_{i\perp}\right) \rho^N_{A(pn)}\left(\alpha^\prime_j,p^\prime_{j\perp}\right)  d\alpha_id^2p_{j\perp}
d\alpha_id^2p_{j\perp},  
\vspace{-0.2cm}
\label{rho3}
\end{eqnarray}
where $(\alpha^\prime_i/j, p^\prime_{i/j\perp})$,  are the LC momentum fractions and transverse momenta of spectator nucleons in the center of mass of $pn$ SRCs.   The nuclear density matrix of $pn$ SRCs are given by  $\rho^N_{A(pn)}(\alpha,p_\perp)$  which, according to $pn$ dominance, are:  $\rho^N_{A(pn)}(\alpha,p_\perp) \approx {a_2(A,Z)\over 2X_N} \rho_d(\alpha,p_\perp)$ where $X_N = Z/A$ or $(A-Z)/A$ is the relative fractions of the proton and neutron in the nucleus and $\rho_d(\alpha,p_\perp)$ is the light-front density function of the deuteron. The factor $F(\alpha^\prime_i,p_{i\perp},\alpha^\prime_j,p_{j\perp})$ accounts for the phase factors of nucleons in the intermediate state between $pn$ interactions and for $0< \alpha_{i/j}^\prime<2$ is a smooth function.

\begin{figure}[htb]
\vspace{-0.2cm}
         \centerline{\includegraphics[width=9cm,height=6cm]{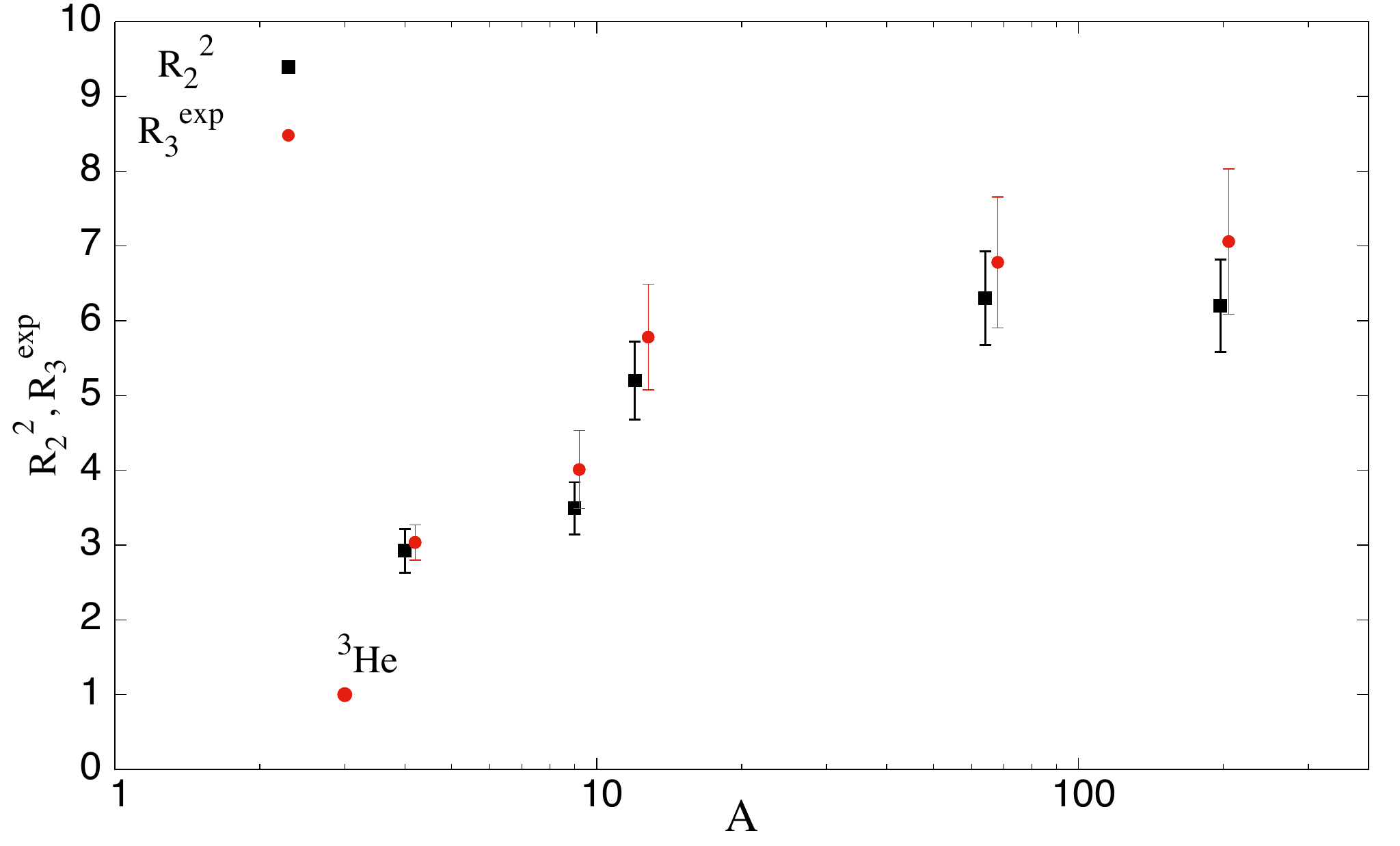} }
         \vspace{-0.4cm}
            \caption{The comparison  of our $R_2^2$ estimates along with the weighted average  in the range $ 1.6 \geq \alpha_{3N} \leq 1.8$ of the ratio data for $A=4,9,12, 64$ and $197$  
}
\label{r3andr2_2}
   \end{figure}

From Eq.(\ref{rho3}) it follows that  the strength of  3N-SRCs is $\sim a_2^2(A,Z)$. This is  seen by calculating the ratio $R_3$ in Eq.(\ref{R3}) using the relation~(\ref{eA}) and the conjecture of Eq.(\ref{rho3}), which  leads to\cite{DFSS18}: 
\vspace{-0.2cm}
\begin{equation} 
R_3(A,Z) =   {9\over 8}{(\sigma_{ep}+\sigma_{en})/2\over (2\sigma_{ep} + \sigma_{en})/3} R^2_2(A,Z)  \approx 
R^2_2(A,z) = \left( {a_2(A,Z)\over a_2(^3He)}\right)^2, 
\vspace{-0.2cm}
\label{R3_to_R2sq} 
\end{equation} 
where $\sigma_{ep} \approx 3\sigma_{en}$  in  the considered $Q^2\sim$ 2.7~GeV$^2$ region. As Fig.\ref{4He_3He_ratio} shows the above prediction of $R_3 \approx R_2^2$ is in agreement with the experimental ratios of $^4$He and $^3$He cross sections.  In the figure the ration $R_2$
is  indicated by  horizontal line for  $ 1.3 \leq \alpha_{3N} \leq 1.5$ where the plateau due to 2N-SRCs is observed. Then, to verify the prediction of 
 Eq.~(\ref{R3_to_R2sq}),   lines are  drawn in the range of $\alpha_{3N} = 1.6-1.8$  at the magnitudes of $R_2^2 = (a_2(A)/a_2(^3He)^2$.
The dashed lines reflects the 10\% error we ascribe to  the prediction of  Eq.~(\ref{R3_to_R2sq}).

There is a similar agreement  for other nuclei including  $^9$Be, $^{12}$C, $^{64}$Cu and $^{197}$Au\cite{DFSS18} however the accuracy of data is not enough for a verification  of the  3N-SRC plateaus at $\alpha_{3N}>1.6$ for 
all considered nuclei.
To test  the prediction of Eq.(\ref{R3_to_R2sq}) we evaluated the weighted average of $R_3(A,Z)$ at $\alpha_{3N}> 1.6$ and compared them with the magnitude 
of   $R_2^2 = {a_2^2(A,Z)\over a_2(^{3He})^2}$  in which $a_2(A,Z)$'s are taken from the analysis of Ref.\cite{Fomin2011}. The results are presented in 
Fig.\ref{r3andr2_2}. They show  a reasonable  agreement with the prediction of Eq.(\ref{R3_to_R2sq}) for a wide range of nuclei.   
 
If the observed  agreement in Fig.\ref{r3andr2_2} is truly due to the onset of  3N-SRCs  then further theoretical analysis will allow  to extract the 
 $a_3(A,Z)$ parameters characterizing the 3N - SRC probabilities in the nuclear ground state\cite{DFSS18}.  Also, 
 with better quality data and a wider range of nuclei in $A$ and $Z$ then the evaluation of the parameter $a_3$ as a function of nuclear density and 
 $pn$ asymmetry can provide an important theoretical input for exploration of the dynamics of super dense nuclear matter.

\vspace{-0.4cm}
 \section{Summary and Outlook}

\vspace{-0.2cm}

From  theoretical analysis of  inclusive processes in the SRC domain  we concluded that the dominating mechanism of 3N-SRCs 
processes corresponds to the situation in which the recoil mass of the 2N spectator system is minimal.  This allowed us to derive a kinematic condition for 
which one may expect the onset of 3N-SRCs in inclusive eA processes that will be reflected in the observation of 
a plateau in the ratio of $eA$ to $e\hspace{1pt} ^{3}\hspace{-1pt}$He  cross sections.  The best quality data available for large enough $Q^2$ (Fig.\ref{4He_3He_ratio}) 
indicate a  possible onset of such a plateau at $\alpha_{3N}> 1.6$. This first signatures of 3N-SRCs is reinforced by the 
good agreement with the prediction of a quadratic ($R_3\approx R^2_2$) dependence between the cross section ratios in  3N-SRCs region, 
$R_3$ and the same ratio  measured in the 2N-SRC region, $R_2$.   Future experiments covering larger values of $Q^2$ will be able to 
verify the above observations unambiguously.  In this respect it is essential that inclusive reactions will be measured at $Q^2\ge 5$~GeV$^2$ which will 
allow to reach the region of $\alpha_{3N}>2$ where one expects clear dominance of 3N SRCs.

{\bf Acknowledgements:}
I am thankful  to Profs. Donal Day, Leonid Frankfurt and Mark Strikman  for collaboration in this work.  The research is supported by the US Department of Energy grant DE-FG02-01ER41172.

\end{document}